\documentstyle[11pt,newpasp,twoside,epsf]{article}

\begin{document}

\title{Pulsational instability domain of $\delta$~Scuti variables}

\author{A.~A.~Pamyatnykh}

\affil{N. Copernicus Astronomical Center, Polish Academy of Sciences,\\
Bartycka~18, 00--716~Warszawa, Poland\\
\smallskip
Institute of Astronomy, Russian Academy of Sciences,\\
Pyatnitskaya~48, 109017~Moscow, Russia\\
\smallskip
Institute of Astronomy, University of Vienna,\\ 
T\"urkenschanzstr.~17, A--1180~Wien, Austria}

\begin{abstract}

An updated theoretical instability domain of the $\delta$~Scuti
star models in the Hertzsprung-Russell diagram, in the $\log g -
\log T_{\rm eff}$ diagram and in diagrams for dereddened $uvby\beta$
photometric indices is presented.  The sensitivity of both the position
of the evolutionary tracks and the Blue Edge of the instability domain to
changes in the chemical composition parameters ($X, Z$) and to changes in
the convection theory parameters (mixing-length in the stellar envelope,
the extent of the overshooting from the convective core) is discussed.

\end{abstract}



\section{Introduction}

This paper continues the study of the pulsational instability domains
in the upper main sequence (Pamyatnykh 1999, Paper~I hereafter).
In Paper~I we considered updated theoretical instability domains in the
Hertzsprung-Russell and in the $\log g - \log T_{\rm eff}$ diagrams for
models of $\beta$~Cephei and SPB stars and studied the influence of
variations of global input parameters (initial chemical composition,
opacity data, efficiency of the overshooting from stellar convective
cores) on the evolutionary models and their oscillations.  Here we
present the results of a similar study for models of $\delta$~Scuti stars.
Our main goal is to demonstrate the influence of the choice of
global parameters on the position of the hotter border (Blue Edge) of
the $\delta$~Scuti instability domain. Linear nonadiabatic oscillation
analyses using the simplest assumption about the interaction between
convection and pulsation (namely, an assumption about frozen-in convective
energy flux during an oscillation cycle) do not allow us to determine
the position of the cooler instability edge (Red Edge). Therefore, in
all diagrams we use an empirical Red Edge following Rodriguez {et al.}
(1994) and Breger (1979) (note that M.\,Breger in his review in these
Proceedings gives a similar but slightly steeper Red Edge). Recent successful
theoretical predictions of the return to stability at the cool boundary of the
instability strip are presented by G.\,Houdek in these Proceedings. 
Perturbations
of the turbulent pressure of convection were found to be the main contributor
to the damping of the pulsations.

There have been other general studies of the $\delta$~Scuti instability
domain.  We note the very important work by Stellingwerf (1979),
who explained the different morphology of the blue edges calculated
for different overtones of radial pulsations. We confirm his results
qualitatively in Section~3. Li \& Stix (1994) and Marconi \& Palla (1998)
used the OPAL opacity data in their computations of the position of
the instability domain. In a paper about slowly pulsating B-type stars,
Dziembowski, Moskalik \& Pamyatnykh (1993) presented an HR diagram where the
Blue Edge 
of the $\delta$~Scuti instability domain was also shown (Fig.\,6 of
the cited paper). Their results were obtained with an earlier version
of the OPAL opacities. We also mention our old results on the blue
edges of the instability domain computed with the Los Alamos opacities
(Pamyatnykh~1975). The first general reviews on $\delta$~Scuti variables,
in which theoretical aspects of the pulsations were outlined, were published
by Baglin {et al.} (1973) and  Breger (1979).

\subsection{Models and their oscillations}

The models of 1.3--3.0 $M_{\odot}$ stars on the main sequence (MS)
and in post-MS evolutionary stages were constructed using a standard
stellar evolution code which was developed in its original version
by B. Paczy\'nski, R. Sienkiewicz and M. Koz{\l}owski (private
communication).  The same code was used in Paper~I and in our recent
seismological studies on individual variables and on period changes in
$\delta$~Scuti stars (see Breger {et al.} 1999 and references therein).
We used the most recent versions of the OPAL and OP opacities (Iglesias
\& Rogers 1996 and Seaton 1996, respectively), supplemented with
the low--temperature data of Alexander \& Ferguson (1994).  In all
computations the OPAL equation of state was used (Rogers et~al. 1996).
The nuclear reaction rates are the same as used by Bahcall and
Pinsonneault (1995).

In one series of the computations the effects of uniform (solid-body)
stellar rotation were taken into account, assuming that the star conserves
its global angular momentum during evolution from the Zero-Age Main
Sequence.

In another series of the computations the possibility of overshooting
from the convective core was taken into account.

The input parameters for evolutionary model sequences are total mass,
$M$, initial values for hydrogen abundance, $X$, and the heavy
element abundance, $Z$.  The initial heavy element mixture is that
of Grevesse \& Noels (1993).  The computations were performed starting
from chemically uniform models on the Zero-Age Main Sequence (ZAMS).
In the stellar envelope, the standard mixing-length theory of convection
with the mixing-length parameter $\alpha=1.0$ was used.

To test the effect of the mixing-length parameter choice on the model
structure and stability,  we computed one family of the models assuming
pure radiative energy transfer in the stellar envelope (i.e., $\alpha=0$).

As in Paper~I, we studied only low-degree oscillations ($\ell \le 2$),
which are those most suitable for photometric detection; the excitation
and visibility of high degree modes in $\delta$ Scuti and other variables
is considered in detail by Balona \& Dziembowski (1999).  A linear
nonadiabatic analysis of the oscillations was performed using a code
developed by W.~Dziembowski (for a general description see Dziembowski
1977).  The effects of slow rotation on the oscillation frequencies were
treated up to third order in the rotational velocity (Dziembowski \&
Goode 1992, Soufi {et al.} 1998)\footnote
{As it is stressed by M.-J.\,Goupil {et al.} in these Proceedings,
rotation is assumed slow enough so that
it may be treated as a perturbation but fast enough so that higher effects
beyond linear in the rotation rate are considered.
}.

\bigskip

In the next section an analogy between $\delta$~Scuti and $\beta$~Cephei
oscillations is demonstrated.  Sect.\,3 contains our main results on
the $\delta$~Scuti instability domain assuming a standard evolutionary
treatment of nonrotating stellar models without overshooting.
The comparison with observational data is also given in this section
using various diagrams.  
In Sect.\,4 we discuss the effects of stellar rotation, convective
overshooting and the mixing-length parameter choice, and in Sect.\,5 we
examine the effect of variations in the chemical composition. 
Last section contains summary and also the discussion of the effect 
of using another set of stellar opacities, namely, OP opacities (Seaton 1996). 
Moreover, the problems of the study of the post-MS $\delta$~Scuti stars are 
outlined using data on the variable 4~Canum~Venaticorum as an example.
Much more detailed results of asteroseismological studies of individual
$\delta$~Scuti stars are presented in the paper 
by J.\,A.\,Guzik et al. in these Proceedings. 

\section{An analogy between the $\delta$~Scuti and $\beta$~Cephei variables}

In the HR diagram, the $\delta$~Scuti stars are located in the lower
part of the classical instability strip where stellar pulsations are
excited by the well-known $\kappa$--mechanism acting in zones of partial
ionization of hydrogen and helium (Baker \& Kippenhahn 1962, 1965; see
also Cox 1963, Zhevakin 1963 and references therein). Chevalier (1971)
was the first who explicitly studied pulsations of a $\delta$~Scuti
star model and concluded that they are excited by the $\kappa$--mechanism
acting in the second helium ionization zone.

The oscillations of the $\delta$~Scuti stars are similar in many
aspects to those of the $\beta$~Cephei variables because in both cases
low-order acoustic and gravity modes are excited by the same classical
$\kappa$--mechanism. The main difference is that the oscillations of the
$\beta$~Cephei 
variables rely on a different opacity bump than those of the stars in the
classical instability strip, where the $\delta$~Scuti variables
are located.  Also, the periods of both groups are similar (lines
of constant period are approximately parallel to the main sequence,
see Fig.\,5 below). We note in this respect (see a review by M.\,Breger
in these Proceedings), that the prototype of the $\delta$~Scuti
group, $\delta$~Scuti itself, was placed initially among the
$\beta$~Canis~Majoris ($\beta$~Cephei) variables according to its period.

Some properties of the pulsations within the $\beta$~Cephei and $\delta$~Scuti 
instability domains are demonstrated in Fig.\,1, where the frequency
oscillation spectra for stellar models of 12 $M_\odot$ and 1.8 $M_\odot$ 
during their evolution from the ZAMS to the TAMS are plotted. 
\begin{figure}[t!]
  \plotone{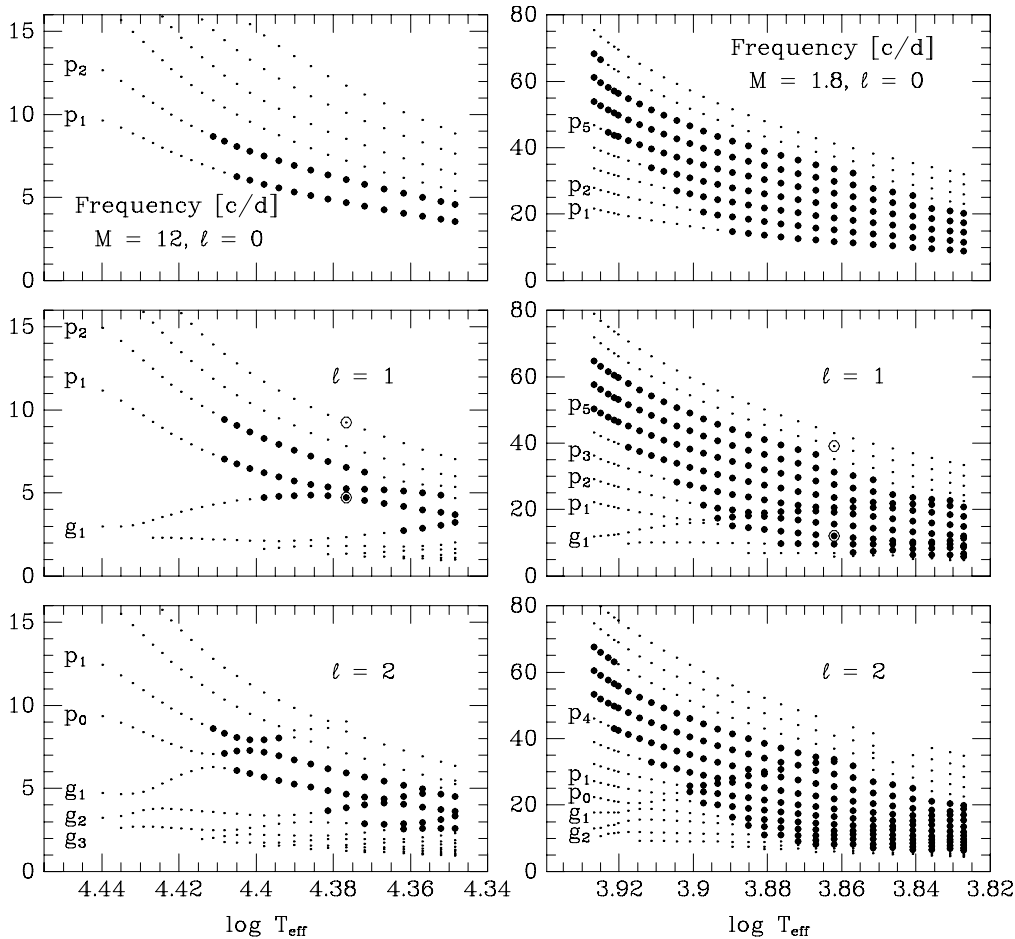}
  \caption[]{
  Frequencies of low-order $p$- and $g$-modes with low degree, $\ell$, for
  models of 12 and 1.8 $M_\odot$ in the Main Sequence evolutionary phase.
  In each panel, the leftmost and rightmost points correspond to the ZAMS
  and TAMS models, respectively.  The large dots mark unstable modes.
  The large open circles in the $\ell=1$ panels mark modes which are shown
  in Fig.\,2.
  }
\end{figure}
For radial modes ($\ell=0$) we see an almost equidistant frequency
separation between consecutive modes.  The complicated patterns of
nonradial modes are caused by evolutionary changes in the stellar
interiors, in the region surrounding the convective core. Due to these
changes, the $p$- and $g$-modes are not separated in frequency already
in mid- or early-MS evolution, and the phenomenon of ``avoided crossing''
between $p$- and $g$- modes takes place (Aizenman {et al.} 1977). This
results in a mixed character of the low-order nonradial modes: they are
similar to pure acoustic modes in the outer stellar layers and to pure
gravity modes in the interiors. The high sensitivity of the avoided
crossing occurrence to the extent of the convective core can be used
to test the efficiency of overshooting from stellar convective cores, as
was proposed by Dziembowski \& Pamyatnykh (1991).

Both in the $\beta$~Cephei and $\delta$~Scuti star models we find
unstable low-order, low-degree $p$-, $g$- and mixed modes.  There is
at least one piece of observational evidence for the presence of $g$- or mixed
modes in the frequency spectrum of the $\delta$~Scuti--type star FG~Virginis
(Breger {et al.} 1999).

We can see that the occurrence of instability during evolution away from
the ZAMS does not depend essentially on mode degree, $\ell$.  Rather,
it is determined primarily by the mode frequency.
Therefore, when discussing the position of the $\delta$~Scuti instability
domain, we restrict ourselves to the study of radial oscillations only.

We find that instability first appears in the acoustic overtones and
then extends to gravity and/or mixed modes.
As can also be seen from Fig.\,1, the frequency range of the unstable
modes in the 1.8 $M_\odot$ models is more extended than that in the 12
$M_\odot$ models.  This is in agreement with the fact that the observed
frequency range of the unstable modes is wider in $\delta$~Scuti than in
$\beta$~Cephei stars. An outstanding example is again FG Virginis, where 23
definitely detected modes span a frequency range of \mbox{9--34~c/d}. 6
or 7 radial overtones are located in this frequency range (see Breger
{et al.} 1999).

For an oscillation mode to be excited by the $\kappa$--mechanism, two
conditions must be fulfilled in addition to the presence of a local
opacity maximum: 
(i) the amplitude of oscillation must be relatively large and slowly
varying in the potentially driving region,
(ii) the thermal timescale in the driving zone, 
$\tau_{_{\mathrm th}}(r) =  {\int\limits_{r}^{R}Tc_{_P}dM}\,/\,L$\,,
must by comparable or longer than the oscillation period. 

If (ii) is not satisfied, the potentially driving region remains in
thermal equilibrium during the pulsation cycle (neutral stability). This
means that in order to excite the oscillations, the opacity bump has to
be located at an optimal geometrical depth in the stellar envelope.

The efficiency of the $\kappa$--mechanism in representative models of
$\beta$~Cephei and $\delta$~Scuti variables is demonstrated in Fig.\,2 (this
figure is a subsection of Fig.\,2 from Pamyatnykh 1999).  The stars are
located in the middle of the corresponding instability domains in the
HR diagram.  Both models have an initial chemical composition of $X =
0.70$ and $Z = 0.02$.  The masses of the models are 12 and 1.8 $M_\odot$,
and the effective temperatures are 23800\,K and 7280\,K, respectively.
All quantities are plotted as functions of temperature inside the models.
\begin{figure}[t!]
  \plotone{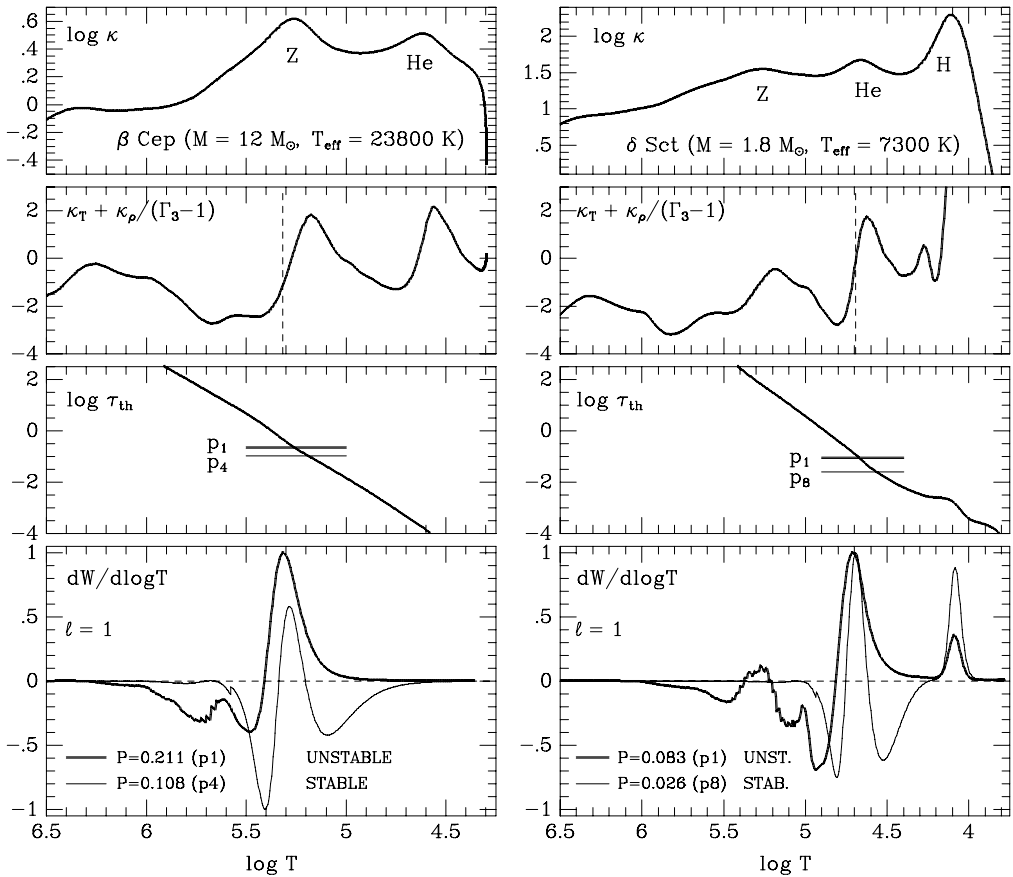}
  \caption[]{
  Opacity, $\kappa$, opacity derivative,
  $\kappa_{_T} + \kappa_{\rho}/(\Gamma_3-1)$,
  thermal timescale, $\tau_{_{\rm th}}$ (in days), and differential
  work integral, $dW/d\log\, T$ (arbitrary units, positive in driving zones),
  for selected pulsation modes ($\ell$=1), plotted versus temperature
  for representative models of $\beta$~Cephei (left) 
  and $\delta$~Scuti (right) variables.
  The dashed vertical lines mark the position of the maximum driving for the
  unstable modes shown in the lower panels.
  The horizontal lines in the $\tau_{_{\rm th}}$ diagrams
  correspond to the periods of selected modes.
  }
\end{figure}
Note that the value of the Rosseland mean opacity, $\kappa$, is systematically
larger for smaller stellar mass, which reflects larger densities
in the envelopes of these stars.

It is easy to see that the main driving in both models takes place in the
layers with a steep radial gradient of the opacity derivative near the
relevant opacity bump: it is the metal or $Z$ bump for the $\beta$~Cephei model
and the bump due to the second helium ionization for the  $\delta$~Scuti
model.

The acoustic mode $p_1$ (as well as other low-order radial and nonradial modes)
is unstable both in the $\beta$~Cephei model and in the $\delta$~Scuti model.
The following dipole ($\ell = 1$) modes are unstable in these
representative models: low-order acoustic modes $p_1 - p_3$ with
periods of 0.211 to 0.153 days in the $\beta$~Cephei model, low-order modes
$g_2$, $g_1$, $p_1 - p_6$ with periods of 0.104 to 0.052 days in the
$\delta$~Scuti model.

The helium bump region, where pulsations of $\delta$~Scuti stars
are driven, does not contribute to the work integral
in the $\beta$~Cephei model due to the very short thermal timescale there.
However, the  timescale requirement can be satisfied in B stars at the
position of the metal bump.

In the model of a $\beta$~Cephei star of 12$M_\odot$, 
the thermal timescale there
is comparable with periods of the low order $p$-modes. The higher-order
acoustic overtones are stable because the damping above the metal bump
region is activated due to their shorter periods (mode $p_4$ in Fig.\,2).
The longer period gravity modes are not excited both due to the
timescale requirement in the metal bump region and due to stronger
damping below this region.

In the model of a $\delta$~Scuti star, the mode $p_1$ (with a period of about 2
hours) is unstable due to the helium opacity bump which is connected with the
second helium ionization zone.  The mode $p_8$  is stable because due to
its three times shorter period the damping between helium and hydrogen
ionization zones is activated.  Note that due to the very sharp hydrogen
opacity bump there is also a small driving contribution from this region.
The metal opacity bump is relatively small in this model. Moreover,
the oscillation amplitudes are small in deep layers. Therefore, the metal
bump (and hence metal abundance) has only a weak influence on stability
of the $\delta$~Scuti models. We confirm this conclusion later showing
only a small sensitivity of the position of the $\delta$~Scuti instability
domain to the choice of initial metal abundance.

\section{Instability domain for a standard set of stellar parameters}

\subsection {Theoretical HR and the $\log g - \log T_{\rm eff}$ diagrams.}

In Fig.\,3 we present the theoretical $\delta$~Scuti instability domain
as computed with the latest version of the OPAL opacities (Iglesias \&
Rogers 1996).  The models were computed without taking into account the
effects of rotation and convective overshooting.  An initial hydrogen
abundance of $X=0.70$ and a heavy element abundance of $Z=0.02$ were
assumed.
\begin{figure}[t!]
  \setlength{\textwidth}{120mm}
  \plotone{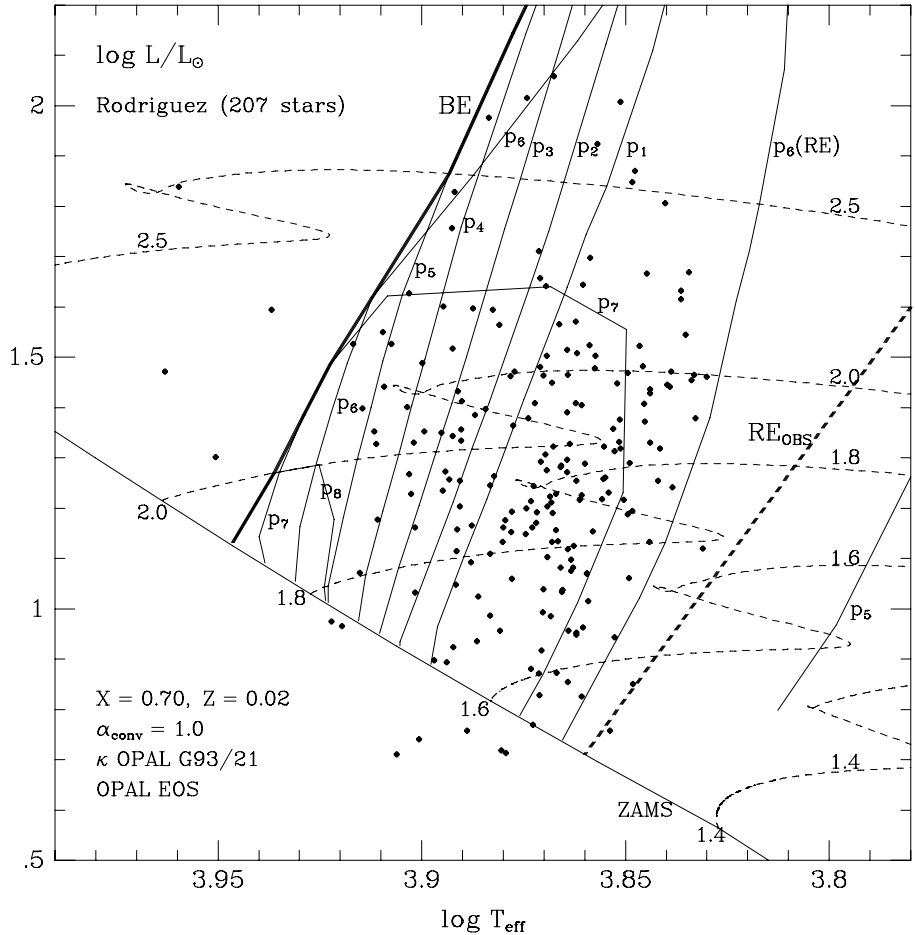}
  \caption[]{
  Theoretical Blue Edges of the $\delta$~Scuti instability domain
  for radial oscillations (symbols $p_1$ - $p_8$ mark corresponding
  modes, starting from fundamental one).  The symbol ${\rm BE}$ stands
  for the general Blue Edge, which is the hotter envelope of unstable
  models.  The symbol ${\rm RE}_{\rm obs}$ marks the empirical Red Edge.
  The Zero-Age Main Sequence (ZAMS) and a few evolutionary tracks for the
  indicated values of $M / M_{\odot}$ are shown.  Observational data are
  from the catalogue of Rodriguez {et al.} (1994).  See text for details.
  }
\end{figure}

Near the ZAMS, the fundamental radial mode, $p_1$, and seven consecutive
overtones, $p_2$ to $p_8$, can be excited. Note that the blue edge
for the radial fundamental mode lies approximately in the center of
the instability strip, far away from the general Blue Edge defined
as the hotter envelope of unstable stellar models.  The Blue Edges
of the unstable regions are hotter for higher overtones. This can be
easily understood if we consider the behaviour of the eigenfunctions
(in particular, the fractional amplitude of the radial displacement)
inside the star and also compare the period of a given mode with the
thermal timescale in the main driving region. Near the Blue Edge of higher
overtones, the fundamental mode and lower overtones are stable both due
to larger amplitudes in the interior (which results in a
relatively stronger damping just below the main driving region) and due to
longer periods (which also results in a stronger influence of the deeper
layers below the driving region if we use the thermal timescale argument
for the instability occurrence, as it was discussed in Section~2).

For cooler stellar models at a fixed luminosity (or along an evolutionary
track), the periods of oscillation are even longer. But the local thermal
timescale in the second helium ionization zone is increasing even faster,
which results in an decreasing influence of the underlying dissipative layers.
Therefore, the lower overtones and the fundamental mode can be exited
as well.

The modes $p_9$ and higher are stable because the damping region above
second helium ionization zone is activated due to their short periods
(the thermal timescale argument) and because of steeper gradients in their
eigenfunctions, which result in smaller amplitudes in the potentially
driving zone of second helium ionization.

Another important property of the blue edges is their curve for higher
overtones.  The symbol $p_6$(RE) in Fig.\,3 marks a ``Red Edge'' for the
mode $p_6$ produced due to a change in the slope of the Blue Edge for this
mode in the same way as for the modes $p_7$ and $p_8$. A similar effect
of intersection of the fundamental and first overtone blue edges with
the luminosity increase is well-known for the RR~Lyrae star models. Cox,
Castor \& King (1972) explained such an intersection by the approach of
a node in the radial displacement to the driving region, which causes a
``quenching'' of the first harmonic instability. The separation of the
main driving region, the second helium ionization zone, from the node in
radial displacement is determined primarily by the ratio $M/R$ of the
global stellar parameters.  Our results confirm those of Stellingwerf
(1979) who concluded that the stabilization of the overtone pulsations
(fourth and fifth in his study) is caused by the approach of the
outermost node of the radial displacement to the driving region.

The Blue Edges computed in this way agree very well with our previous
results based on earlier OPAL opacities (see Fig.\,6 in Dziembowski,
Moskalik \& Pamyatnykh 1993, where the general Blue Edge of the $\delta$~Scuti
instability domain is plotted together with instability domains of the
SPB and $\beta$~Cephei stars).  There is also good agreement with our old
results obtained with Los Alamos opacities for the fundamental radial mode
and third overtone (Pamyatnykh 1975): old BE\,($p_1$) lies close to BE
($p_2$), and Blue Edges for the third overtone, $p_4$, are close to
one another. This quite satisfactory agreement between results obtained
with the OPAL and Los Alamos data are explained by the fact that in
the most important region around the second helium ionization zone both series
of opacities do not differ significantly; the main difference is confined
to much deeper layers (the metal opacity bump at $T \sim 200\,000$\,K is
missing in the Los Alamos data) which have practically no effect on the
$\delta$~Scuti-type instability near the Blue Edges.

Our blue edges BE(p1) and BE(p3) agree quite well 
with those by Stellingwerf (1979) (ours are slightly steeper). 
Note that Stellingwerf considered only stellar envelope models.

Blue edges of the $\delta$~Scuti domain computed by Li \& Stix (1994) are
significantly hotter. However, these results disagree
with data about XX~Pyxidis, whose lowest unstable mode frequency is that of
mode $p_4$ 
or $p_5$ (see Pamyatnykh {et al.}\,1998).
To achieve agreement with the Li \& Stix instability results, the star must
be evolved and must have a mass of about 
2.5 $M_{\odot}$. However, calibrated photometric data give $\log g \approx
4.25$ which means that the star is located close to the ZAMS and its mass is 
about 1.9 $M_{\odot}$. 

To compare the position of the theoretical domain with the observations, we
chose 207 stars with available $uvby\beta$ photometry from the catalogue
of Rodriguez {et al.} (1994).  Note that in the general HR diagram
for the upper main-sequence stars (see Fig.\,3 in Paper~I, which is
reproduced  in Breger's review in these Proceedings) the observed
135 $\delta$~Scuti variables from the catalogue of Garcia et al. (1995) are
plotted. 
The photometric data were transformed to effective temperature,
$T_{\rm eff}$, using the calibration due to Moon \& Dworetsky (1985)
(in practice, we used the program UVBYBETA written by Moon 1985 and
modified by R.\,Napiwotzki).  Stellar absolute magnitudes, $M_V$, were
derived also with this program; the calculations are based on the relations
of Str\"omgren (1966) and Crawford (1975, 1979), with some modifications.
Finally, bolometric corrections to the absolute magnitudes were determined
according to the calibration by Balona (1994).

Our comparison with the observations is rather illustrative. We do not
discuss here the errors of calibrations, the positions of individual
stars and we do not try to correct some data using additional information
(like distances for the members of open clusters or according to the
HIPPARCOS data). Our goal is to demonstrate the possibility of using
different diagrams for such a comparison and to note some regularities.

As can be seen from Fig.\,3, there is a quite good general agreement
of the theoretical and observational domains of instability. Also, some
regularities can be seen in this diagram.  Approximately 25\% of the
variables are hotter than the Blue Edge for the radial fundamental mode,
$p_1$, so they must pulsate only in overtones.  Also, many variables,
approximately 30\% of the total number, are located above the main
sequence band, whose boundary is determined by the Terminal-Age Main
Sequence (the TAMS line is not plotted in this figure; it connects the
rightmost points of the MS part of each evolutionary track). The situation
differs from that of $\beta$~Cephei stars (see Paper~I), nearly all of
which are located on the main sequence. The explanation is very simple:
the post-MS evolution of massive stars is very fast in comparison with
the MS evolution, therefore we do not observe $\beta$~Cephei variables during
this stage. On the other hand, in the $\delta$~Scuti domain, the post-MS
evolution is only 10 to 50 times faster than the MS evolution (see Breger
\& Pamyatnykh 1998).  Moreover, selection effects (higher luminosities
and statistically higher amplitudes of the post-MS variables) favour 
detecting the post-MS $\delta$~Scuti variables. Also, overshooting from the
stellar convective core (see Sect.\,4), if it exists, will result 
in a broadening of the main sequence band and thereby will influence
the percentages of the MS and post-MS pulsators. In the future, we plan to
study this problem in more detail.

A few stars are hotter than the general Blue Edge, as was noted 
by Rodriguez {et al.} (1994), who called for their systematic observations in
order to compare with the theoretical predictions in detail 
(see also Schutt~1993). A standard photometric calibration may be wrong for
these stars, if they are chemically peculiar. Also, the theoretical Blue Edge
is hotter for higher helium abundance, as will be demonstrated in Section~5.

Moreover, a few stars are located under the ZAMS. Probably, they are
chemically peculiar, as was discussed by Rodriguez {et al.} (1994).
If this is the case, then a standard $M_V$ calibration should not be
applied to these stars.

In Fig.\,4 the $\delta$~Scuti instability domain in
the $\log g - \log T_{\rm eff}$ diagram is shown.
The same models as in Fig.\,3 were used.
The photometric data were transformed to $T_{\rm eff}$ and $\log g$, using
the calibration due to Moon \& Dworetsky (1985) (as for Fig.\,3, we used the
program UVBYBETA previously referred to).

\begin{figure}
  \setlength{\textwidth}{120mm}
  \plotone{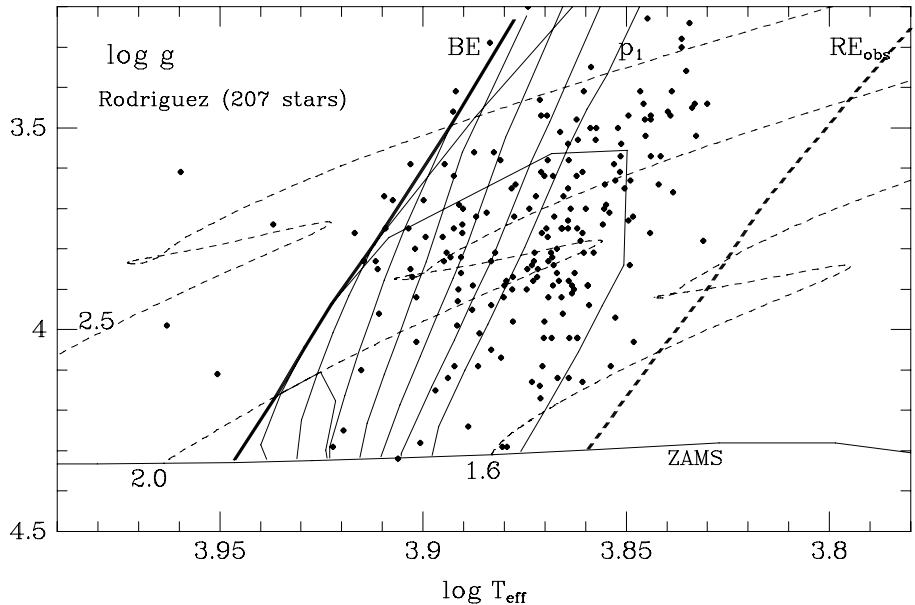}
  \caption[]{
  The $\delta$~Scuti instability domain in
  the $\log g - \log T_{\rm eff}$ diagram.
  The same models and the same calibrations for observed stars as in Fig.\,3
  were used. 
  A few evolutionary tracks for the indicated values
  of $M / M_{\odot}$ are shown.
  }
\end{figure}

If we  compare the theoretical and observed instability domains, such a diagram
seems to be more useful than the HR diagram because here we use only one
calibration based on models of atmospheres. For the HR diagram (and for
the $M_V - (b-y)_0$ diagram in Fig.\,6 below), we used a
separate calibration of the absolute magnitudes which is based on other data.

As for the HR diagram in Fig.\,3, we see here a good agreement between theory
and observations; most of the variables are located within the theoretical
instability domain.

However, approximately 50\%  of the variables seem to be in a post-MS
evolutionary stage in this diagram, in comparison with 30\% for the HR
diagram in Fig.\,3.  This difference may be caused by relatively fast
rotation, which decreases the effective gravity due to centrifugal
acceleration.

It is easy to estimate that in the region around the TAMS the centrifugal
acceleration may reach about 20\% of the gravity at $V_{\rm rot}
\approx 150$~km/sec and about 40\% at $V_{\rm rot} \approx 200$~km/sec,
which results in a decrease in $\log g$ of 0.1 -- 0.2. The shift of the
star positions in Fig.\,4 by 0.1 to higher $\log g$ values will result
in a decrease of post-MS pulsators to approximately 30\% of the total
number, in agreement with our estimate for the HR diagram in Fig.\,3.
Moreover, the stars which are slightly hotter than the Blue Edge will
be located within the theoretical instability domain after such a shift
in gravity, also in agreement with the HR diagram in Fig.\,3.

In Fig.\,5 lines of constant period for the radial fundamental mode in
the $\log g - \log T_{\rm eff}$ diagram are shown.  The same models as
in Fig.\,3 and 4 were used. Points mark the position of all evolutionary
models of $1.5-2.5 M_{\odot}$ used to obtain lines of constant periods
by interpolation with the IDL graphics package.
\begin{figure}
  \setlength{\textwidth}{130mm}
  \plotone{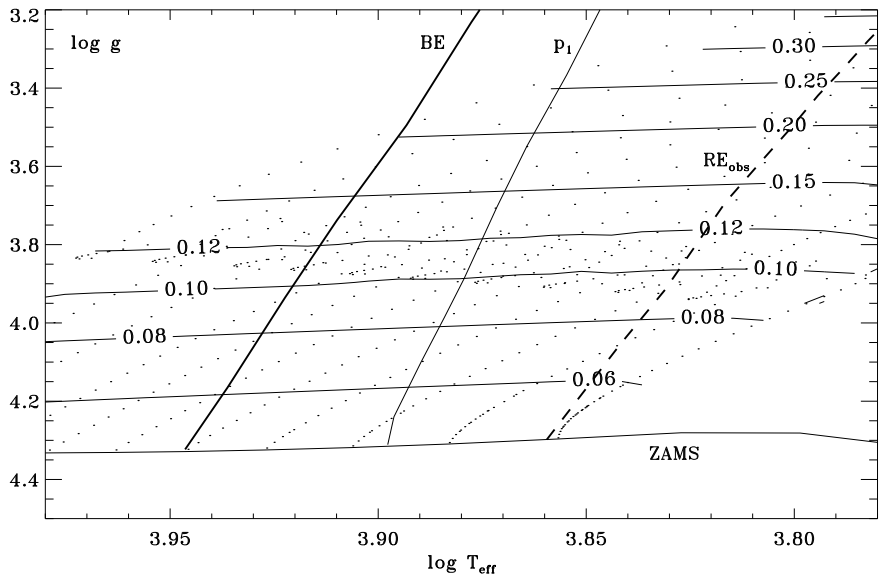}
  \caption[]{
  Lines of constant period for the radial fundamental mode in
  the $\log g - \log T_{\rm eff}$ diagram.
  The same models as in Fig.\,4 were used.
  Period values are given in days.
  }
\end{figure}
Using this plot and the calibrated $uvby\beta$ photometric data it
possible immediately to compare the observed period with the theoretical
period of the radial fundamental mode for a model of normal chemical
composition.  Note that for the models of $\delta$~Scuti stars periods 
of the first,
second and third radial overtones are approximately 77\%, 62--64\%
and 52--54\% of the fundamental period, respectively.

\subsection{Instability domain in photometric coordinates}

When comparing observational and theoretical data on instability domains,
we can use an inverse approach in contrast to the transformation of the
observed photometric indices to the theoretical parameters.  Namely, we
can try to transform the theoretical parameters, which determine the position
of the models in the HR and $\log g - \log T_{\rm eff}$ diagrams, to
the observable photometric indices. To construct such diagrams, we used
theoretical $uvby\beta$ indices which were computed by Lester, Gray \&
Kurucz (1986) using different Kurucz model atmospheres.

Note that the photometric calibrations of effective temperature and
gravity of Moon \& Dworetsky (1985) and of Lester {et al.} (1986) differ
from one another.  Fortunately, for relatively cool A and F stars with
$T_{\rm eff} \la 8500$\,K  (practically all $\delta$~Scuti variables
belong to this group) both calibrations give effective temperatures
in good agreement with stellar temperature standards, as discussed by
Napiwotzki {et al.} (1993a).

So, our diagrams in photometric coordinates will differ from previous
theoretical diagrams not only in the coordinates themselves, 
but to some extent by using different calibrations.

Fig.\,6 shows the $\delta$~Scuti instability domain in
the $M_V - (b-y)_0$ diagram. The theoretical data from Fig.\,3
(ZAMS line, evolutionary tracks and blue edges of the pulsational instability
domain for different radial overtones) 
have been transformed to photometric indices according to  
the Lester, Gray \& Kurucz (1986) tables for unpublished Kurucz grid for solar
composition. Cubic spline interpolation in $T_{\rm eff}$ and $\log g$ has been
used to obtain the theoretical $uvby\beta$ indices.
The empirical Red Edge (${\rm RE}_{\rm obs}$) according 
to Rodriguez {et al.} (1994) and Breger (1979) is also shown
(note that the new empirical RE given by Breger in his review 
in these Proceedings is slightly steeper than the old one).

Different symbols have the same sense
as in Fig.\,3.
\begin{figure}[t!]
  \setlength{\textwidth}{120mm}
  \plotone{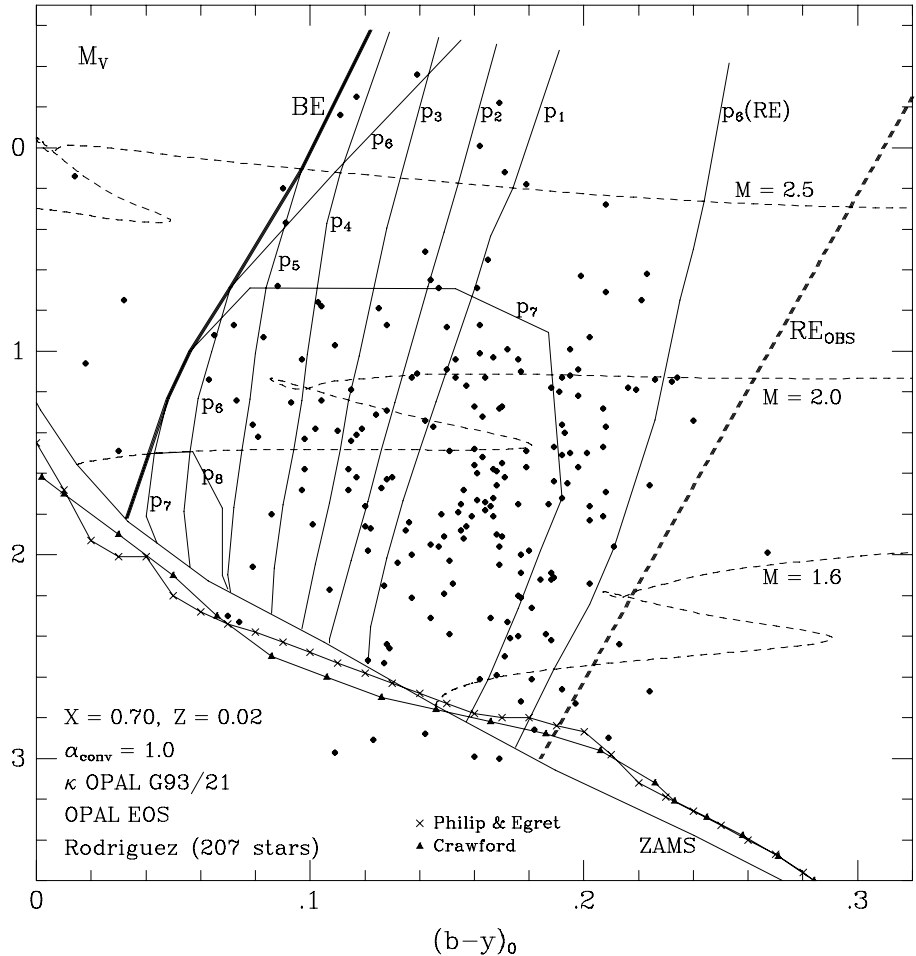}
  \caption[]{
  $\delta$~Scuti instability domain in
  the $M_V - (b-y)_0$ diagram. 
  Empirical ZAMS lines (i.e., reference lines or unreddened standard relations
  between parameters) according to Crawford (1975, 1978, 1979) and 
  Philip \& Egret (1980) are marked by triangles and crosses, respectively.
  }
\end{figure}
The observational photometric data from the Rodriguez {et al.} (1994)
catalogue have been dereddened according to the procedures of Crawford
(1979) as they are implemented in the program UVBYBETA (Napiwotzki {et
al.} 1993a).

Absolute magnitudes, $M_V$, were calculated also with this program using
the relations  of Str\"omgren (1966) and 
Crawford (1975, 1979) with some modifications. 

For the stars with $T_{\rm eff} \la 8500$\,K 
($\log T_{\rm eff} \la 3.93$), the value of $\beta$, a measure of the equivalent
width of the ${\rm H}_{\beta}$ line, is a good temperature indicator and $c_0$,
a measure of the strength of the Balmer discontinuity, is a good gravity
parameter (see, for example, Napiwotzki {et al.} 1993a). 
Also, the color index $(b-y)_0$ is a good temperature indicator.
Practically all
$\delta$~Scuti variables belong to this group of relatively cool stars.

We can see that there are systematic differences between the empirical and
the theoretical ZAMS. The existence of this problem is demonstrated also
by Breger for different calibrations in terms of stellar mass (see Fig.\,7
of his review in these Proceedings). In addition to uncertainties in the
photometric calibrations, we note that the position of the theoretical
ZAMS is very sensitive to the choice of global stellar parameters, mainly
to the initial chemical composition, as will be shown in Section~5.

There is a quite good agreement between theoretical and 
observational instability domains, with the same general regularities as in
the theoretical HR diagram (Fig.\,3).

Figs.\,7 and 8 show the $\delta$~Scuti instability domain in the $(c_1)_0
- (b-y)_0$ and in the $(c_1)_0 - \beta$ diagrams, respectively.  As in
Fig.\,6, the theoretical data have been transformed to photometric indices
according to the Lester, Gray \& Kurucz (1986) tables for unpublished
Kurucz grids for solar composition.  The observational photometric data
from the Rodriguez {et al.} (1994) catalogue have been dereddened using
the UVBYBETA program.

\begin{figure}[t!]
  \setlength{\textwidth}{120mm}
  \plotone{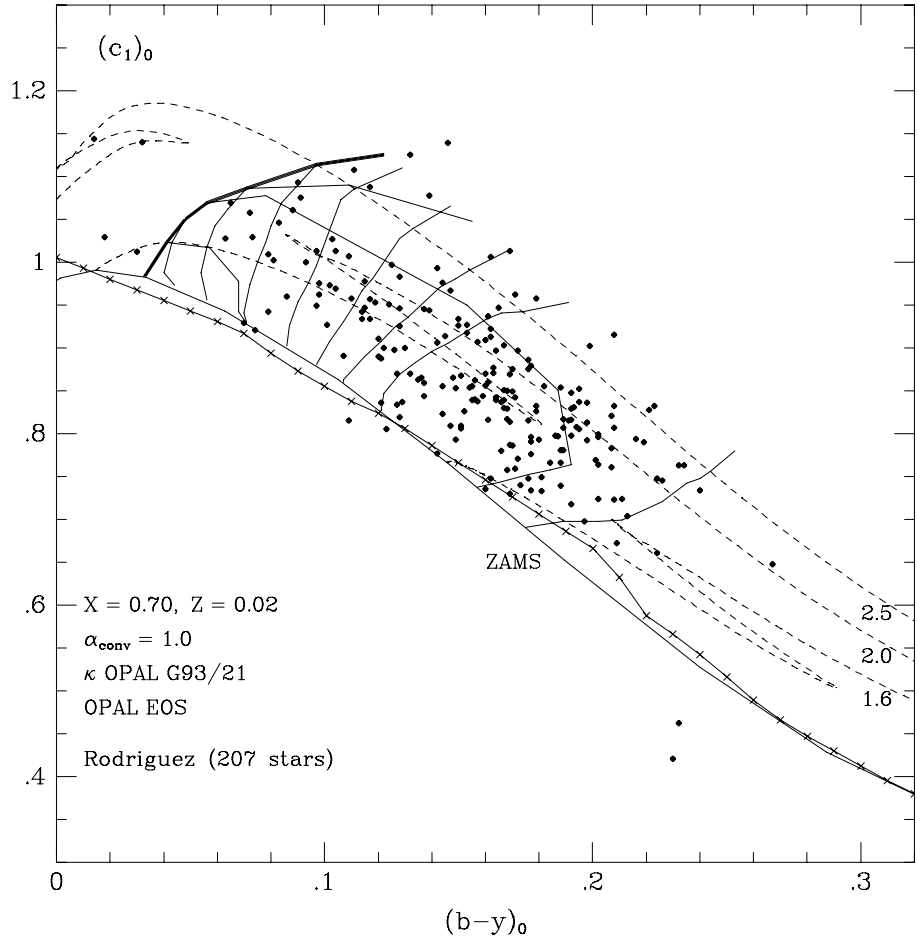}
  \caption[]{
  The $\delta$~Scuti instability domain together with the observational
  data from the Rodriguez et al. (1994) catalogue  in the $(c_1)_0 -
  (b-y)_0$ diagram.  The empirical reference line according to Philip \&
  Egret (1980) is marked by crosses.
  }
\end{figure}
Two stars below the ZAMS are the chemically peculiar variables AU Scl (Am-type)
and XZ Men ($\delta$~Del-type). 
(Both in the $M_V - (b-y)_0$ diagram and in the theoretical diagrams
of Figs.\,3 and 4 they are located outside the
parameter ranges of the figures). The luminosities of these stars are
underestimated significantly (and gravities are overestimated) if we use
a standard calibration of $uvby\beta$ photometry which is not valid 
for chemically peculiar stars (see discussion in Rodriguez {et al.} 1994).

The systematic difference between the theoretical and the empirical ZAMS
in Fig.\,7 is qualitatively similar to that in Fig.\,6. In our opinion,
this similarity can be considered as an indication of the consistency of
two separate photometric calibrations---the $M_V$ calibration and the
model atmosphere calibration used to obtain $\log g$ and  $\log T_{\rm
eff}$ values.  The last calibration was used to transform all theoretical
lines from $\log g$, $\log T_{\rm eff}$ coordinates to the photometric
indices in Figs.\,7 and 8.  This difference between the theoretical
and the empirical ZAMS is even more pronounced in Fig.\,8. Here, we
only demonstrate the effect which must be taken into consideration in
detailed comparisons of observed data with theoretical models.

\begin{figure}
  \setlength{\textwidth}{120mm}
  \plotone{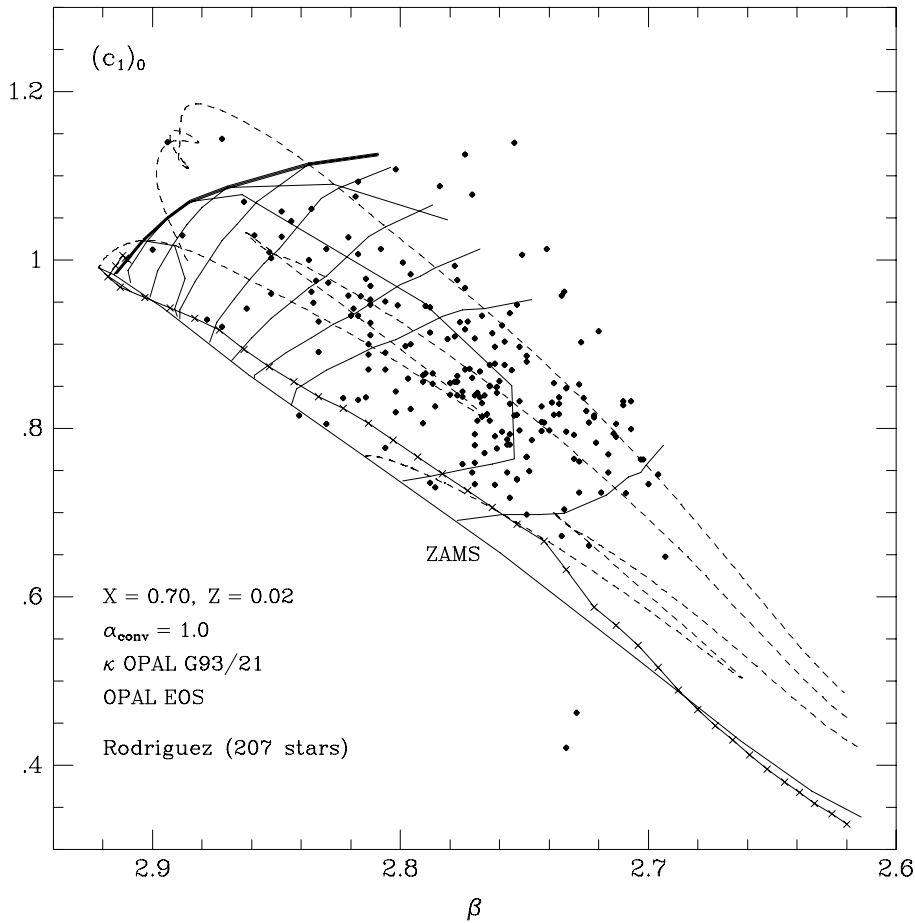}
  \caption[]{
  The $\delta$~Scuti instability domain together with the observational
  data from the Rodriguez et al. (1994) catalogue  in the $(c_1)_0 -
  \beta$ diagram.  The empirical reference line according to Philip \&
  Egret (1980) is marked by crosses.
  }
\end{figure}

It can be seen from Fig.\,8 that a lot of variables are located beyond
the main sequence band (i.e., beyond the TAMS, which can be imaged as
a line which connects the rightmost points of zigzag parts of different
evolutionary tracks,
much more than in the $M_V - (b-y)_0$ diagram (Fig.\,6). A similar
effect was noted already in the $\log g - \log T_{\rm eff}$ diagram as
compared with the HR diagram (Figs.\,4 and 3, respectively). It may be
caused by stellar rotation which results in decreasing the effective
gravity on the stellar surface and therefore changes the photometric
index $c_1$.  Indeed, Crawford (1979) found a correlation between the
$\delta c_1$ parameter (which is a measure of the distance from ZAMS)
and the rotational velocity for the Pleiades and $\alpha$\,Persei stars:
the $\delta c_1$ index, which is sensitive to gravity, indicates a lower
gravity for the more rapid rotators.

Crawford also noted that this effect may not be significant for most F stars
which are rotating slower than A stars. 

Lester {et al.} (1986) discussed the effect of rotation on the
$\beta$~index due to rotational broadening of the ${\rm H}_{\beta}$
line and due to the change of its equivalent width as a result of the
rotational changes of the interior and the atmosphere structure. For
a main-sequence model with $T_{\rm eff} = 9500$~K the $\beta$~index
decreases by only 0.005 mag as the rotational velocity increases from
0 to 300 km/s.

Some stars in Fig.\,8 are located to the right and to the top of most
of the variables. From 6 stars here with known values of $v\,\sin\,i$,
4 stars have values between 140 and 190 km/sec.

\section{Effects of convection, convective overshooting and stellar rotation}

\subsection{Effect of convection}

To estimate the influence of the efficiency of convection in stellar
envelopes on the position of the instability domain we computed a family
of models completely neglecting convective energy transfer in the envelope
(however, nothing was changed in the treatment of the convective stellar core)
and studied their stability. The results of this test are presented
in Fig.\,9.
\begin{figure}[t!]
  \setlength{\textwidth}{120mm}
  \plotone{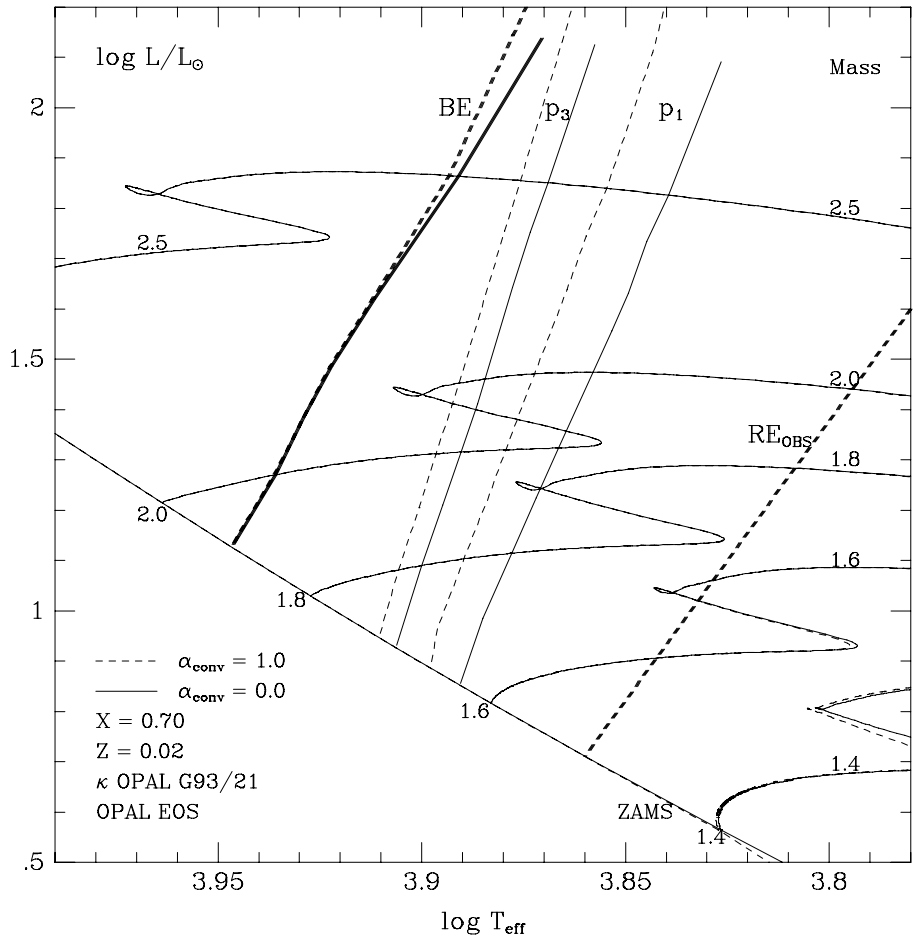}
  \caption[]{
  The effect of neglecting convection in the stellar envelope on the
  position of the blue edges of the $\delta$~Scuti instability domain.
  The Zero-Age Main Sequence (ZAMS) and a few evolutionary tracks for
  the indicated values of $M / M_{\odot}$ are shown.
  }
\end{figure}

We can see that in the $\delta$~Scuti instability region convection in the
envelope has no influence on the tracks and ZAMS position. A noticeable
effect takes place only for the models with $M \la 1.4 M_{\odot}$.

Also, convection does not influence the position of the general Blue
Edge. However, it changes the position of the radial fundamental Blue
edge: BE($p_1$) is approximately $200-250$\,K  cooler if convection is
artificially suppressed in the stellar envelope. Such behavior is caused
mainly by structural differences between convective and pure radiative
models in the region of the hydrogen ionization which contributes to
the driving of pulsations as well as in the main driving region of the
second helium ionization. In our computations we neglect the Lagrangian
variation of the convective energy flux during an oscillation cycle,
i.e., we assume that the convective flux is at all times equal to the
unperturbed value. This assumption is used generally (see, for example,
Baker \& Kippenhahn 1965). G.\,Houdek in these Proceedings discusses a
more reliable theory for describing the interaction between convection
and pulsation, taking into account dynamical effects of convection.

The convective energy flux is negligible in models near the general
Blue Edge.  But near the Blue Edge of the fundamental mode,  convection
can transfer up to 80--90\,\% of the energy in the center of the hydrogen
ionization zone at temperatures of about 11000\,K and therefore influence
the structure of this zone. On the red edge, the convective contribution
to the energy flux in the hydrogen ionization zone may grow as large as
99\,\%. Note that the efficiency of convection is negligible in the second
helium ionization zone in all models within the $\delta$~Scuti instability
domain. For example, in a 2\,$M_{\odot}$ model the convective contribution
does not exceed 0.5\,\% near the fundamental radial blue edge and 5\,\%
near the red edge.

\subsection{The effect of overshooting from the stellar convective core}

Some indications in favor of significant overshooting from the convective
cores of $\delta$ Scuti stars were obtained from a comparison of
evolutionary tracks with calibrated photometric data (e.g., Napiwotzki
{et al.} 1993b).  An asteroseismological test, based on the sensitivity
of some nonradial mode frequencies to the size of the mixed stellar core,
was proposed by Dziembowski \& Pamyatnykh (1991).

To test the effects of overshooting explicitly, we computed a family of
evolutionary tracks of 1.4--3.0\,$M_{\odot}$ for models with overshooting
from the convective core and studied the stability of these models.
The results are presented in Fig.\,10. This figure is similar to that
of Breger \& Pamyatnykh (1998) where the Blue Edges are presented only
for the standard models without overshooting.

The overshooting distance, $d_{\rm over}$, was chosen to be $0.2\,
H_{\rm p}$, where $H_{\rm p}$ is the local pressure scale height at
the edge of the convective core. A similar value of the overshooting
parameter was used by many authors (see, for example, Schaller {et al.}
1992, Napiwotzki {et al.} 1993b, Claret 1995).
\begin{figure}
  \setlength{\textwidth}{130mm}
  \plotone{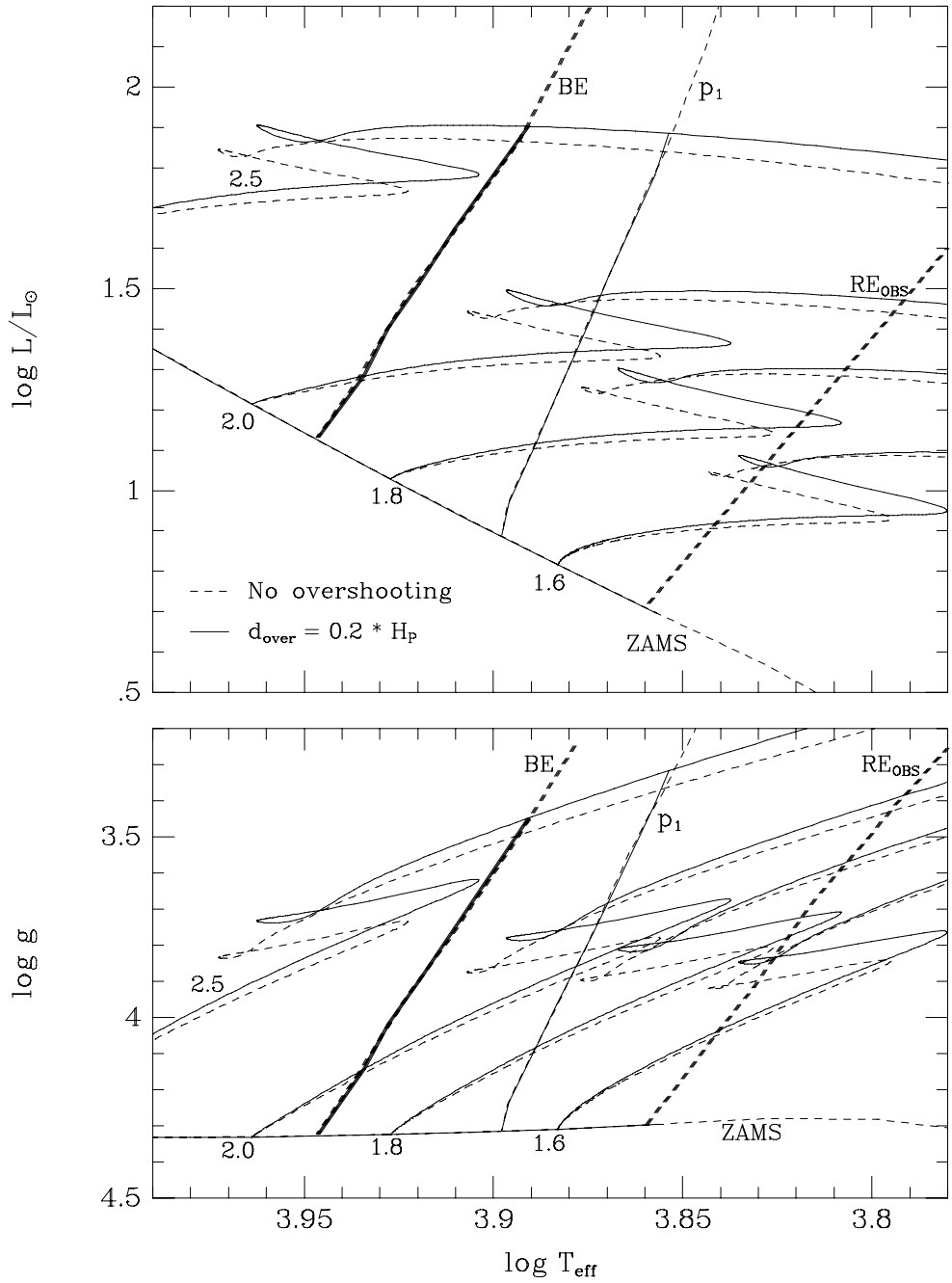}
  \caption[]{
  The effect of overshooting from the convective stellar core on the
  evolution and on the position of the blue edges of the $\delta$~Scuti
  instability domain in the HR and in the $\log g - \log T_{\rm eff}$
  diagrams.  The Zero-Age Main Sequence (ZAMS) and a few evolutionary
  tracks for the indicated values of $M / M_{\odot}$ are shown.
  }
\end{figure}

The ZAMS models are identical in both cases (with and without
overshooting) because they are chemically uniform. 

The overshooting results in an extension of the MS-stage in the HR
diagram due to an enlargement of the mixed core: more hydrogen fuel is
available for nuclear burning.
The displacement of the TAMS-points is about
$\Delta (\log T_{\rm eff}) = -0.02$,
$\Delta (\log g) = -0.1$ to $-0.15$\footnote
{
According to our computations with a given choice
of $d_{\rm over}$, overshooting influences the TAMS position 1.5--2
times less than according to Schaller {et al.} (1992), Napiwotzki
{et al.} (1993) or Claret (1995). The disagreement may be caused by
numerical effects as a test shows. To achieve a sufficient accuracy
in the oscillation computations we need to rely upon more detailed
stellar models than one uses usually in the evolutionary computations.
For example, our typical 2.0 $M_{\odot}$ MS model consists of
approximately 1300 layers, and there are 135 models between the ZAMS
and TAMS. If we increase both space and time step sizes by factor
of about 7--8, i.e., use more crude space and time grids, 
we can reproduce the corresponding evolutionary track of
Schaller {et al.} (1992) almost precisely.
}.
At a fixed effective temperature, a model with overshooting is
slightly more luminous than a standard model without overshooting.
Stellar lifetimes in the MS-stage are increased due to overshooting
by 12--14\%.  At the TAMS, hydrogen is slightly more exhausted in models
with overshooting: the central hydrogen abundance is about 4.0\% (in mass)
as compared with 4.5\% in the case without overshooting.

It is interesting to compare stellar lifetimes in different evolutionary
stages for the models with and without overshooting.  Let us consider
segments of the evolutionary tracks of the 1.8 $M_{\odot}$ model in the
effective temperature range between the TAMS and the leftmost point
of the second contraction stage. The star crosses that region three
times in its life on and beyond the MS. In the overshooting case ($\log
T_{\rm eff}=3.8081-3.8671$, see Fig.\,10) these times are equal to 0.35,
0.034, and 0.019 Gyr for MS, second contraction and post-MS expansion
stages, correspondingly.  In the standard case without overshooting
($\log T_{\rm eff}=3.8260-3.8769$) these times are equal to 0.35, 0.038,
and 0.056 Gyr. We see that in both cases the second contraction times
are similar and are one order of magnitude shorter than in the final
part of the MS evolution.  On the other hand, post-MS expansion in the
overshooting case is a factor of 1.8 faster, and in the standard case a
factor of 1.5 slower than the corresponding second contraction. Such a
significant difference in the post-MS expansion times between evolution
with and without overshooting seems to be important for statistical
investigations of the disribution of the $\delta$~Scuti stars in the HR
and similar diagrams.

Fig.\,10 shows very clearly that overshooting from the convective core
does not affect the position of the instability domain at all: the Blue
Edges for the cases with and without overshooting coincide because the
stellar envelope structure in both cases is very similar.

(In the same way, there must be no significant influence on the instability
domain if we consider, for example, pre-MS stars which also differ mainly
in their interior structure from MS stars. Parameters of the instability strip
for pre-MS models have been computed by Marconi \& Palla (1998). Our Blue Edge
for the second overtone, $p_3$, is hotter than that of Marconi \& Palla
by approximately 0.02 in $\log T_{\rm eff}$. Such a difference needs to be
explained in more detailed comparison of both computations which is beyond
the scope of the present paper.)

Note that our studies of the $\beta$~Cephei instability domain
(Dziembowski \& Pamyatnykh 1993, Pamyatnykh 1999) showed that for a normal
chemical composition no overshooting is required to achieve a very good
agreement between observed data and the period range of unstable radial
modes in the main-sequence stellar models in the Period\,--\,$\log T_{\rm
eff}$ diagram. Also in the HR diagram no extension of the MS stage via
convective overshooting is required to fit all $\beta$~Cephei variables
with the MS band quite well.  We cannot test the overshooting theory
in the same way for the $\delta$~Scuti stars because many variables are,
probably, in the post-MS evolutionary stage and therefore a test based
on the extent of the MS band cannot be applied in this case. As it was
noted already (Dziembowski \& Pamyatnykh 1991), there exists another
asteroseismological test of convective overshooting theories but we do
not yet have suitable observational data.

\subsection{Effect of stellar rotation}

An important parameter, which must influence stellar evolution and
pulsations, is rotation. Detailed studies of non-evolutionary rotational
effects on the oscillation frequency spectrum are beyond the scope
of the present review; they are discussed by M.-J.\,Goupil {et al.} in these
Proceedings.  However, rotation significantly modifies the structure
and evolution of a star.

In addition to the study by Goupil {et al.} we present here two more
figures which illustrate the effect of rotation on stellar models and on
the oscillation frequency spectrum.  To this goal, we computed a family
of evolutionary tracks of 1.4--3.0\,$M_{\odot}$ for rotating models.

We assumed uniform (solid-body) stellar
rotation and conservation of global angular momentum during evolution
from the ZAMS. These assumptions were chosen due to their simplicity.
The same approach was used in our Paper~I and in our papers on $\delta$~Scuti
variables (see Breger {et al.} 1999 and references therein).
The initial equatorial rotational velocity was assumed to be 150 km/s.
The tracks for rotating and non-rotating models are shown in Fig.\,11.

\begin{figure}
  \setlength{\textwidth}{130mm}
  \plotone{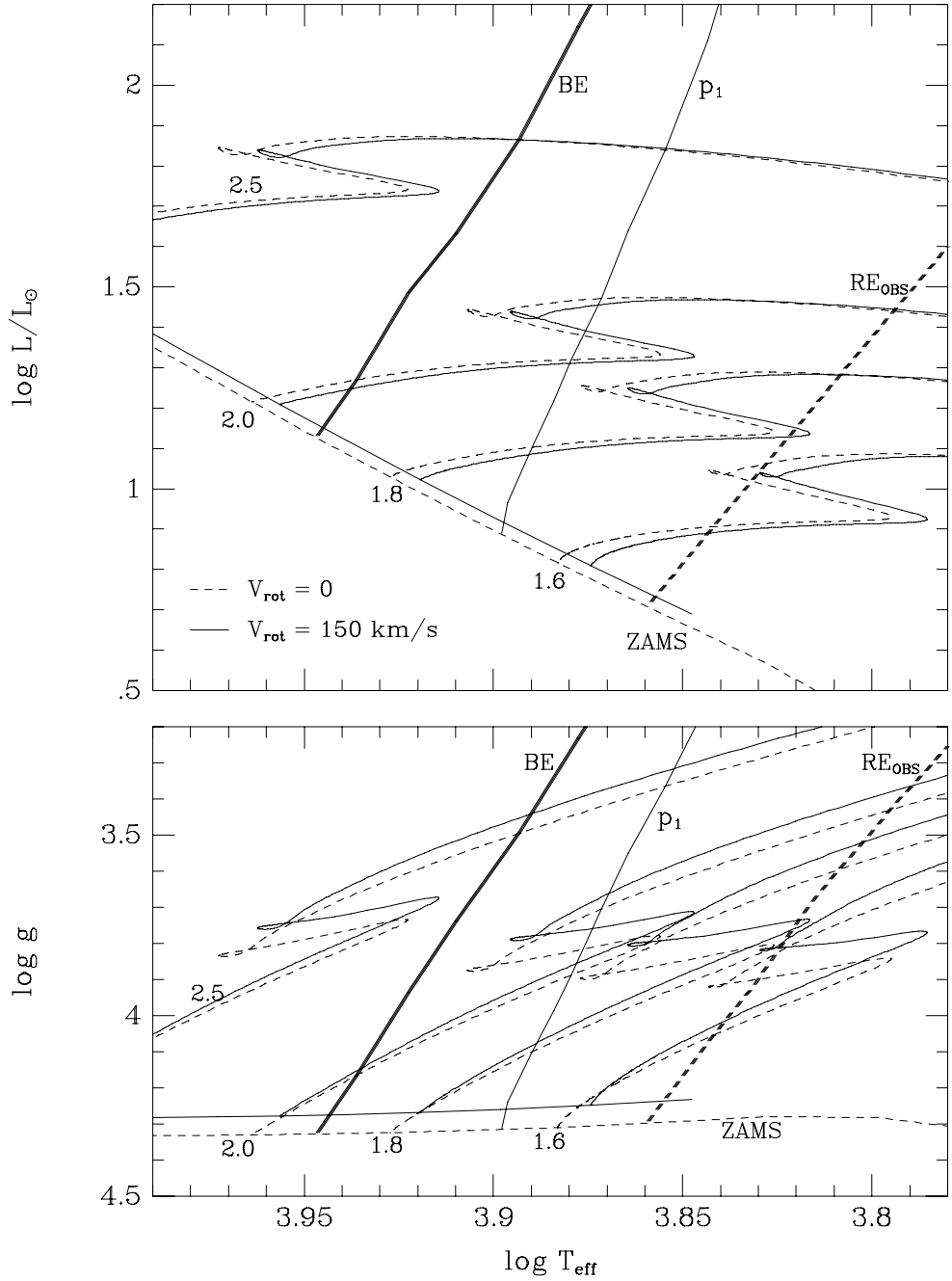}
  \caption[]{
  Evolutionary tracks of uniformly rotating (solid lines) and 
  non-rotating (dashed lines) stellar models 
  with masses 1.6, 1.8, 2.0 and 2.5 $M_{\odot}$. (See text for details.)
  The instability boundaries are the same as in Fig.\,3.
  }
\end{figure}

Rotation results in a shift of the tracks to smaller $T_{\rm eff}$
and gravity. The displacement of the TAMS mimics to some extent the
overshooting effect, with $\Delta (\log T_{\rm eff})$, $\Delta (\log g)$
smaller by a factor 1.5 in comparison with those for the overshooting case
(compare Figs.\,10 and 11).  However, in contrast to the overshooting case,
there is a noticeable shift of the ZAMS, especially in the $\log g -
\log T_{\rm eff}$ diagram.  The tracks of rotating models lie slightly
above the tracks of non-rotating models in the $\log g - \log T_{\rm eff}$
diagram, and below the corresponding tracks in the $\log L - \log T_{\rm
eff}$ diagram which can be easily explained by a decreased effective
gravity due to a horizontally averaged centrifugal force.

The MS lifetime for rotating models is only 0.5--1.0\%
longer than that for non-rotating models.
With our assumption of global angular momentum conservation,
the equatorial rotational velocity decreases during the
MS-evolution from 150 km/s on the ZAMS to about 120 km/s at the TAMS.

As has been noted by Goupil {et al.} in these Proceedings, differential
rotation  is likely to exist in main sequence stars and result in a mixing
of chemical elements in the radiative interior.  The evolutionary tracks for
relevant stellar models are presented in the paper by Goupil {et al.} in
these Proceedings. The mixing in the radiative stellar interiors influences
the position of the track more strongly than a change of the stellar
structure due to centrifugal acceleration.  Therefore to obtain more
realistic estimates of the rotation effects on the oscillation frequency
spectrum and on instability we must take these effects into account.

However, here we present as an illustration the results assuming the
simplest law of uniform rotation, in addition to the detailed discussion
given by Goupil et al. in these Proceedings.

We did not compute yet the blue edges of the instability for rotating
models, therefore only blue edges for the standard nonrotating models
are given in the figure. According to computations by Lee \& Baraffe
(1995), the effects of centrifugal force and the resultant rotational
deformation of the main-sequence stars do not affect significantly the
pulsational instability of $p$- and $g$-modes.

The effects of slow rotation on the
oscillation frequencies were treated up to third order in the rotational
velocity (Dziembowski \& Goode 1992, Soufi {et al.} 1998,
Goupil {et al.} in these Proceedings).

Fig.\,12 demonstrates the change in frequencies of low-degree modes in the
observed frequency range of XX~Pyxidis for models of different rotational
velocities. The radial modes are the overtones $p_4$, $p_5$, $p_6$.
\begin{figure}
  \setlength{\textwidth}{130mm}
  \plotone{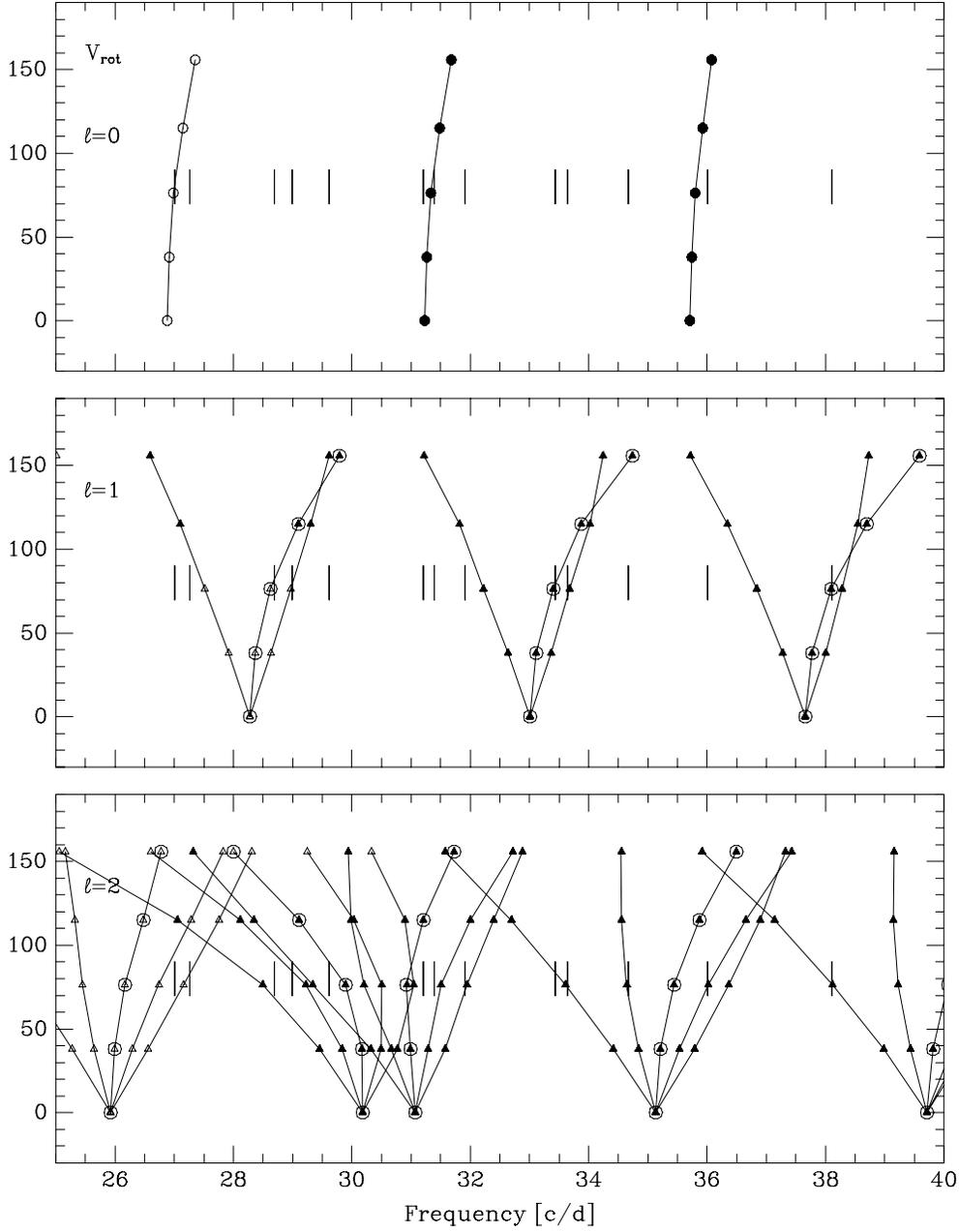}
  \caption[]{
  Rotational splitting of the frequencies of low-degree oscillation
  modes for models of 1.9 $M_{\odot}$ near the Main Sequence.  All models
  of different $V_{\rm rot}$ have the same effective temperature $\log
  T_{\rm eff} = 3.91$. The luminosity of the models varies from $\log
  L = 1.21$ to $\log L = 1.23$ depending on $V_{\rm rot}$ (smaller for
  faster rotation).  Vertical lines mark the observed frequency spectrum
  of XX Pyxidis.
  }
\end{figure}

The non-spherical deformation due to centrifugal force destroys the
symmetry of the rotationally split multiplet; the effect is larger for
higher $p$-mode frequencies.  The asymmetry is already noticeable at a
rotational velocity of about 40--50 km/sec.  The effect of rotational
coupling of nearly degenerate modes, which is not taken into account
in the figure (see Goupil {et al.} in these Proceedings) results in
additional perturbations to the frequency spectrum: close modes  with
degrees $\ell$ differing by an even integer  and with the same $m$ value
are being repelled. This complicates the fitting of the observed and
computed frequency spectra, as was demonstrated by Pamyatnykh {et al.}
(1998) in an attempt to find a seismic solution to the 13 observed
frequencies of XX Pyxidis.

It can also be seen that for $\ell=2$ there are two close modes just
before an avoided crossing. The mode with lower frequency is mainly a
gravity mode, it has smaller asymmetry of the rotational splitting than
the mode with higher frequency.

Finally, we note that in the nonlinear oscillation regime a coupling between
modes within a multiplet may enforce an equidistant splitting, as it is
discussed by Goupil {et al.} in these Proceedings.

\section{Effects of variation of chemical composition}

In Paper I we studied the sensitivity of the position of the $\beta$~Cephei
and SPB instability domains to variations in the initial abundance of
hydrogen, $X$ (or helium, $Y$), and heavy elements, $Z$.  It was shown
that the efficiency of the $\kappa$--mechanism acting in the region of the
metal opacity bump is very sensitive to $Z$ and much less sensitive to
$X$, a result that is not surprising.  For example, with decreasing $Z$
the $\beta$~Cephei instability domain shrinks down. At $Z = 0.01$ there
are no unstable modes in the observed $\beta$~Cephei region (cf.~Fig.\,3
in Paper~I).

Here we show the results of similar tests for the $\delta$~Scuti instability
domain.  It was demonstrated in Section~2 that oscillations
of the $\delta$~Scuti variables are excited mainly in the second helium
ionization zone. Therefore, the efficiency of the $\kappa$--mechanism
is not expected to be sensitive to the heavy element abundance but it
should be strongly affected by the helium (or hydrogen) abundance.

\subsection{Changing metallicity}

In Fig.\,13, we demonstrate the effect of variations in the heavy
element abundance on the position of the blue edges of the $\delta$~Scuti
instability domain. Indeed, the effect is insignificant. Smaller $Z$
means a slightly higher helium abundance, $Y$, because we fixed the hydrogen
abundance, $X$, at a standard value $X = 0.70$. Therefore, the driving
in the second helium ionization zone is slightly more effective.

Note the very high sensitivity of the ZAMS line and evolutionary tracks
to the heavy element abundance. The shift of a track of given mass
to higher effective temperature and higher luminosity, when the $Z$
value is decreasing, can be easily understood in terms of the opacity
because a smaller $Z$ value means a smaller value of the opacity in the
interiors. Very simple and clear examples of the effect of opacity changes
on evolutionary tracks were given almost twenty years ago by Fricke,
Stobie \& Strittmatter (1971).  The width of the MS band decreases only
slightly as $Z$ decreases; the main effect of a $Z$ decrease is a shift
of the the MS as a whole.

\begin{figure}
  \setlength{\textwidth}{130mm}
  \plotone{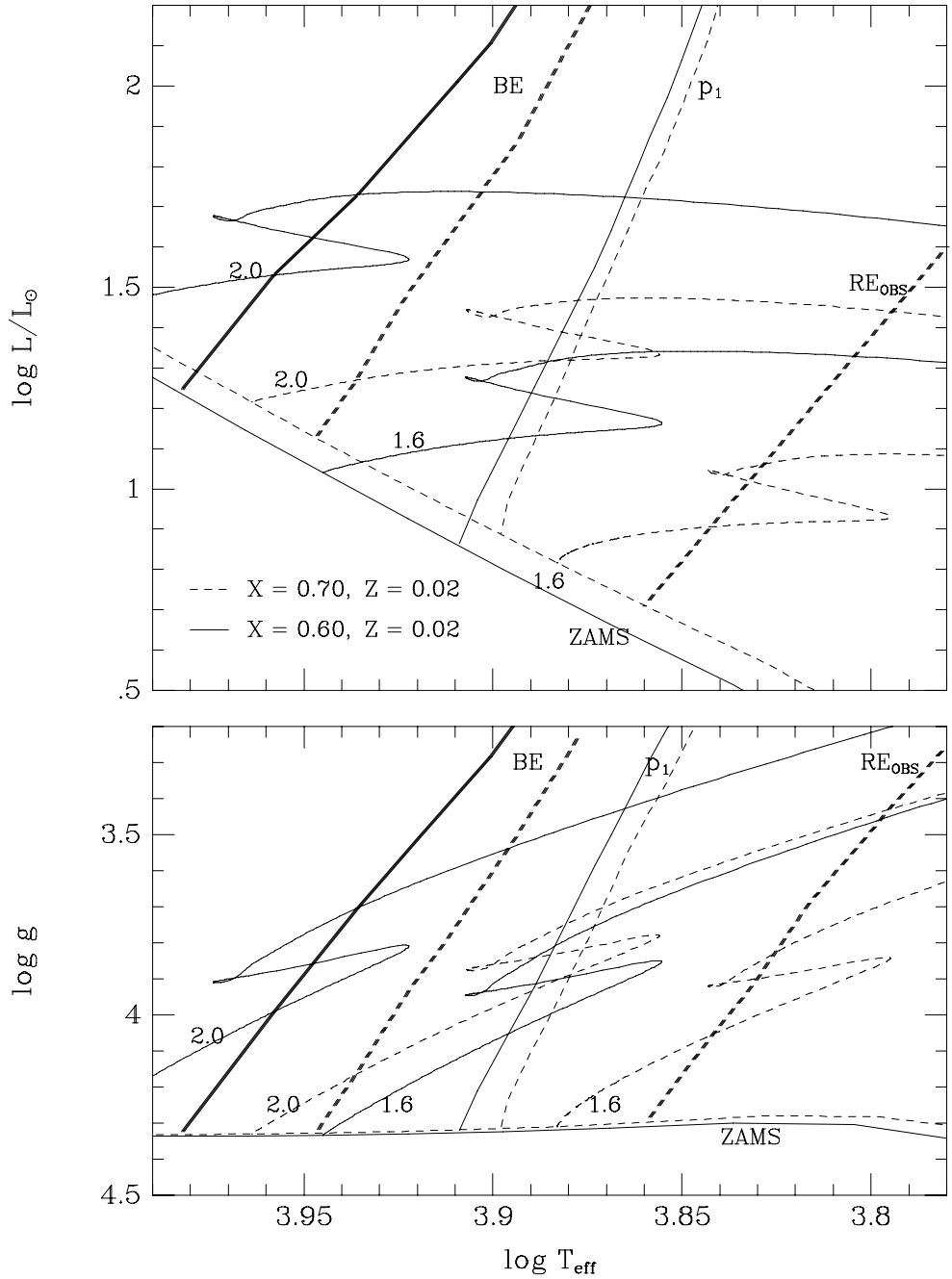}
  \caption[]{
  The effect of changes in the heavy element abundance
  on the evolution and stability of the models of $\delta$~Scuti stars.
  Dashed lines show the results obtained for the standard
  chemical composition $X=0.70, Z=0.02$ (Fig.\,3 and 4). 
  For both $Z=0.01$ and 0.02,
  the ZAMS lines and evolutionary tracks for the two indicated values
  of $M / M_{\odot}$ are shown.
  }
\end{figure}

We would like to stress that there exists a significant uncertainty in
the stellar mass value for a given point in the HR or $\log g - \log
T_{\rm eff}$ diagrams due to uncertainties in chemical composition.
Identified oscillation modes and good photometric calibrations may help
to reveal some information about the chemical composition.

\subsection{Changing initial hydrogen abundance}

In Fig.\,14 we show the effect of the hydrogen abundance change
on the evolution and stability of the $\delta$~Scuti stars.
Note, that in the $\log g - \log T_{\rm eff}$ diagram, 
the ZAMS line practically does not change with $X$ change. The ZAMS model
of a fixed mass shifts along this line when the initial abundance has been
changed. It can be seen also, that
the decrease of the initial hydrogen abundance results in the narrowing
of the MS band (less fuel is available for nuclear burning).

As expected, a change in $X$ influences the position of the Blue Edges,
but the effect appears to be very strong only for the general Blue Edge
and for the blue edges of higher overtones; it becomes smaller for lower
overtones and smallest for the radial fundamental mode. Approximately
25\,\% of the shift of the general Blue Edge near the ZAMS to hotter
effective temperatures is caused by the addition of the unstable overtone
$p_9$, which is not excited for the standard chemical composition.

The relatively small effect of a change in $X$ on the fundamental Blue
Edge position is caused by the fact that the increase of driving in the
second helium ionization zone with increasing helium abundance is partly
compensated by the decrease of driving in the hydrogen ionization zone which
also contributes to the excitation of oscillations for the relatively
cool stellar models near the fundamental radial Blue Edge. For hotter
models near the general Blue Edge the hydrogen ionization zone does not
contribute to the excitation of oscillations.
We stress that the instability or stability of models near blue edges
is determined by a quite small difference between strong driving and strong
damping in relevant stellar layers.

The high sensitivity of the position of the general Blue Edge to the hydrogen
(or helium) abundance can be used to estimate (constrain) these abundances
in observed pulsating stars located close to the blue edges.

\begin{figure}
  \setlength{\textwidth}{130mm}
  \plotone{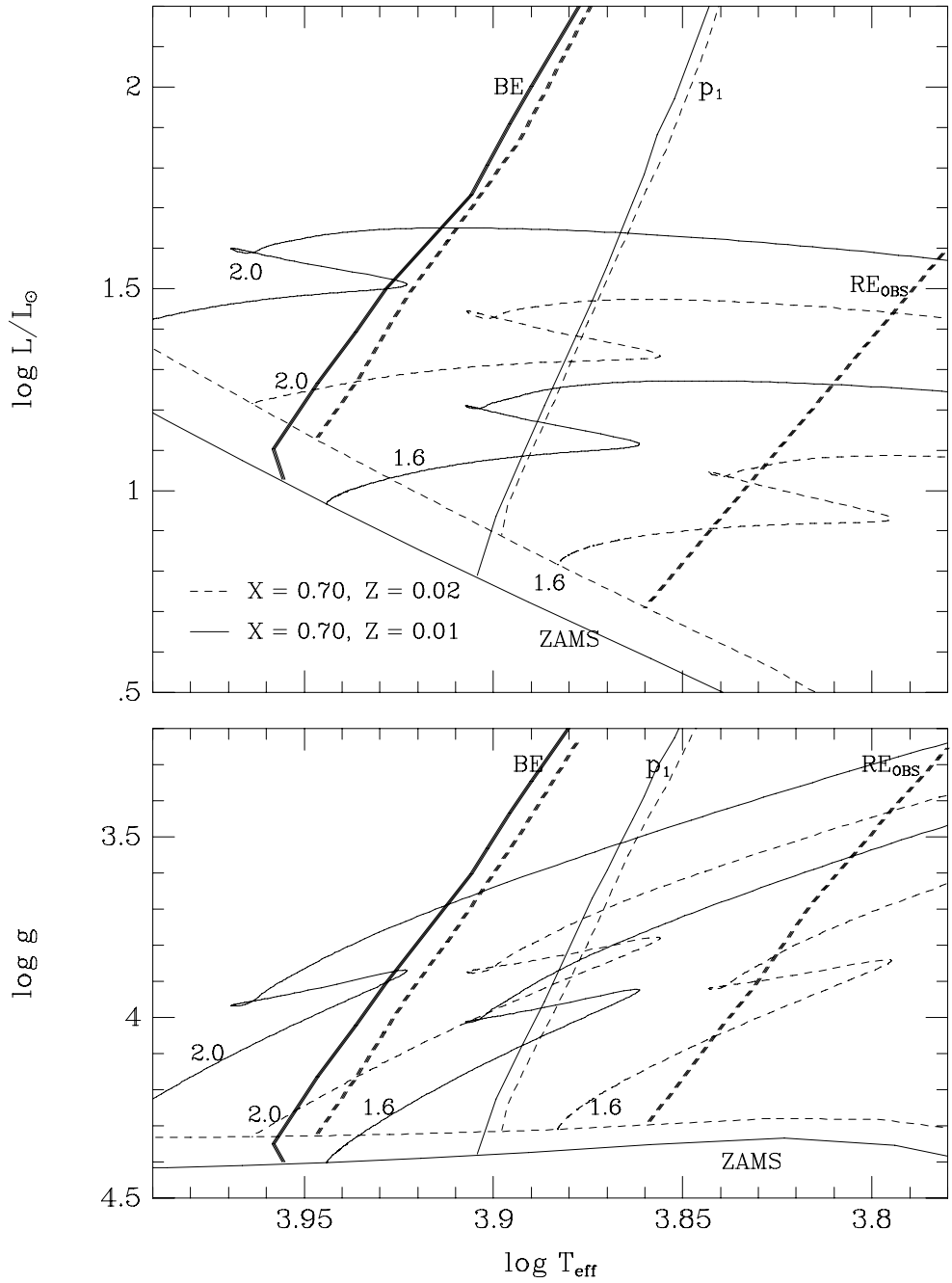}
  \caption[]{
  The effect of the hydrogen abundance change
  on the evolution and stability of the models of $\delta$~Scuti stars.
  Dashed lines show the results for the standard
  chemical composition $X=0.70, Z=0.02$ (Figs.\,3 and 4).
  For both $X=0.60$ and 0.70,
  the ZAMS lines and evolutionary tracks for the two indicated values
  of $M / M_{\odot}$ are shown.
  }
\end{figure}

\section{Discussion and some problems for future studies.}

In this paper we presented the updated theoretical instability domain 
of the models of $\delta$~Scuti stars in various diagrams and studied
the sensitivity of the position of the evolutionary tracks 
and the blue edges of the instability to changes in
the global stellar parameters. In particular, we demonstrated that the
convection theory parameters (mixing-length in the stellar envelope,
the extent of the overshooting from the convective core) and the abundance of
heavy elements do not influence the position of the general Blue Edge of the
instability.  In contrast, 
the high sensitivity of the position of the general Blue Edge to the hydrogen
(or helium) abundance can be used to constrain these abundances
in the observed pulsating stars located close to the blue edges.

We would like to summarize our results concerning the uncertainty of
the ZAMS position due to initial chemical composition variations. Our
computations show that at a fixed effective temperature of $\log T_{\rm
eff} \approx 3.9$, i.e., in the middle of the instability region on the
ZAMS, the ZAMS line for $Z=0.04$ lies above the ZAMS line for $Z=0.01$
by $\Delta \log L \approx 0.28$, and the ZAMS line for $X=0.80$ lies
above the ZAMS line for $X=0.60$ by $\Delta \log L \approx 0.16$. In
the $\log g - \log T_{\rm eff}$ diagram, the corresponding differences
are $\Delta \log g \approx 0.135$ and $\Delta \log g \approx 0.015$,
respectively. Note the practical insensitivity of the ZAMS to initial
hydrogen abundance changes in the $\log g - \log T_{\rm eff}$ diagram
(Fig.\,14); the effect is confined here to the displacement of a given
stellar mass point along the ZAMS. The effect of the variations of the
heavy metal abundances in the $\log g - \log T_{\rm eff}$ diagram is
seen very clearly. Such a difference in the ZAMS shift due to hydrogen
and heavy element variations in the two diagrams can be potentially used
to choose between changes of hydrogen or heavy element abundances when
comparing observational data with the theoretical calculations.

\subsection{On the effect of using the OP opacities}

All results presented in this paper were obtained using the OPAL opacities.
To test the influence of the opacity choice on the instability, we computed a
family of models using another set of opacities, namely, OP
opacities (Seaton 1996).

The two series of opacity data differ from one another both due to
completely independent approaches to calculate stellar opacities and due
to small differences in heavy element abundances: the abundances of Ni
differ by about 5\%, and those of Ne, Cr and Fe by 2-3\%. Note that
all heavy element abundances are close to those for the solar mixture
(Grevesse \& Noels 1993).  The computations with the OP opacities were
performed using the same opacity interpolation routines as used for OPAL.

The OP opacities are systematically lower than OPAL in the deep
stellar interiors.  As was shown by Iglesias \& Rogers (1995),
these significant discrepancies are caused mainly by the fact that
the OP atomic data neglect inner-shell photo-excitations which can
contribute to the opacity. It was also shown by these authors that the
OP approach makes approximations in the implementation of the occupation
probability formalism that reduces the number of electrons in excited
bound states. The combination of decreased excited level populations in
OP plus the missing atomic data reduce the Rosseland mean opacity up to
30\% at high densities and explain the observed discrepancies between
OPAL and OP data.

Breger {et al.} (1999) computed models of a $\delta$~Scuti-type star,
FG~Virginis, 
both with OPAL and OP opacities. For these  models,
the difference in opacity is about 20\% at temperatures above
$10^6$~K. In the envelope, at lower temperatures, the OP opacity varies
slightly more monotonously along radius than does OPAL opacity: some dips
are slightly shallower and some bumps are more flat. The differences
do not exceed 8\%: for example, at a temperature of $14\, 000$~K the OP
opacity is 4\% smaller and at a temperature of $300\, 000$~K it is 7.5\%
higher than the OPAL opacity. 

The significant discrepancies in the very deep stellar interiors do not
affect the instability which is caused by the second helium ionization
zone at a temperature of about 50\,000\,K. We showed in Paper~I that
even the instability due to the metal opacity bump at a temperature of
about 200\,000\,K, which is responsible for the $\beta$~Cephei and SPB star
oscillations, does not depend qualitatively on the choice between OPAL
and OP opacities.  In these driving regions the OPAL and OP data do not
differ significantly.

Therefore, we do not expect any significant differences in the position
of the Blue Edges of the $\delta$~Scuti instability domain when using OP 
instead of OPAL opacities. The explicit computations fully confirm this 
expectation; therefore, we do not present any diagram for the OP case.

Moreover, in the regions of helium and hydrogen ionization in stellar
envelopes, the new OPAL and OP opacities do not differ significantly from
the old Los Alamos opacities. Therefore our new results concerning the
Blue Edges of the instability are in good agreement with earlier results
obtained with Los Alamos data, as was noted in Section~3.

Note, however, that the smaller OP opacities in the deep interiors result
in a shift of the evolutionary tracks to higher effective temperatures
and luminosities, just like the shift of the tracks due a decrease in
the metal abundance (Fig.\,13).

\subsection{Post-MS variables: 4~Canum Venaticorum as an example}

To illustrate the complexity of the oscillation frequency spectra of
evolved post-MS models of $\delta$~Scuti-type stars, we show in Fig.\,15
the oscillation frequency spectrum for a model of 4 Canum Venaticorum
which oscillates with at least 17 frequencies in the range 4.7--9.7 c/d.
The  ordinate gives the oscillatory moment of inertia ($I$) which is
proportional to the oscillation kinetic energy and which is evaluated
assuming the same radial displacement at the surface for each mode.
The high density of the oscillation spectrum of nonradial modes is caused
by very large values of the Brunt-V\"ais\"al\"a frequency in the deep
interior.

The theoretical frequency range of unstable modes agrees very well
with the observed range.  Modes partially trapped in the envelope are
characterized by a low value of $I$.  The trapped modes are analogs
of pure acoustic modes.  It can be seen that the trapping is much more
effective for dipole modes ($\ell = 1$) than for quadrupole modes ($\ell
= 2$).  The total number of unstable modes of $\ell \le 2$ is equal to 449
taking into account the rotational splitting of each mode into $2\ell +
1$ components (only 111 unstable modes of azimuthal order $m = 0$ are
shown in the figure).

So, if the model is correct, less than 4 percent of the unstable modes
are excited to observable amplitudes.  Potentially, the mode selection
mechanism may be related to enchanced mode trapping in the envelope
(Dziembowski \& Kr\'olikowska 1990).  It may be easier to excite them
to a given amplitude on the surface.  However, as it has been noted by
Dziembowski (1997), ``we do not know whether this effect has anything to
do with mode selection.  It is possible that the amplitudes are random
quantities and that there is no rule of mode selection. Without such
a rule we will never be able to make use of nonradial mode frequencies
for seismic probing.''

\begin{figure}[t]
  \setlength{\textwidth}{120mm}
  \plotone{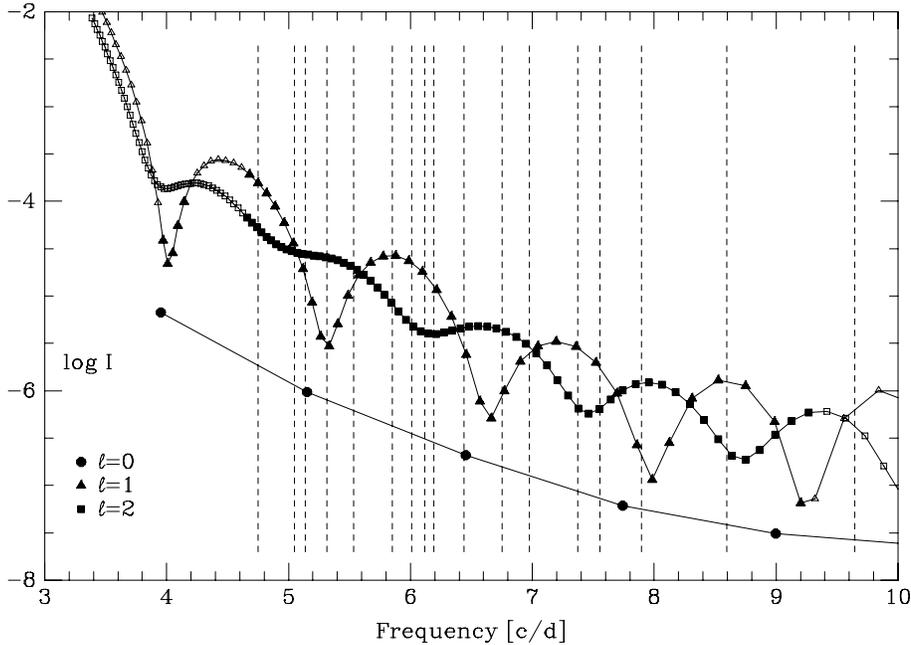}
  \caption[]{
  Oscillation frequencies in 4 Canum Venaticorum (shown as vertical lines)
  compared with those in an approximate model ($M = 2.3 M_{\odot},
  \log T_{\rm eff} = 3.84, V_{\rm rot} = 88$ km/sec, $\log g = 3.39$). 
  The ordinate gives the oscillatory moment of inertia ($I$) which is evaluated
  assuming the same radial displacement at the surface for each mode. 
  Only axesymmetric modes ($m=0$) are shown. 
  Open symbols denote stable modes.
  From Dziembowski (1997).
  }
\end{figure}

In these Proceedings, Breger reports the observed spacing of about 1.2 c/d 
between identified $\ell=1$ modes, but the theoretical spacing between 
trapped \mbox{$\ell=1$} modes in Fig.\,15 is seen to be about 1.4 c/d. 
If we assume that we do observe the trapped modes then,
to decrease the theoretical spacing, it is necessary to
increase the stellar mass and/or to decrease the effective temperature
in order to obtain $\log g \approx 3.3$, which is still within 
the allowed range based on the photometric calibrations.

\subsection{On $\gamma$~Doradus variables}

Quite recently, a new group of variable F stars on the cool side of
the $\delta$~Scuti instability domain was classified, $\gamma$~Doradus
variables (see a paper by A.\,B.\,Kaye in these Proceedings and also Kaye {et
al.} 1999). Typical periods are from 0.4 to 3 days which corresponds
to high-order gravity mode pulsations.  The observational domain of
these variables in the HR diagram is discussed by Handler (1999a). This
instability region partly overlaps with that of the $\delta$~Scuti variables. 
As was noted by Handler (1999a) if stars exhibiting simultaneous
$\gamma$~Dor and $\delta$~Scuti-type pulsations could be found, this
would increase the possibilities for asteroseismology of both classes of
variables. Higher-frequency $p$-mode oscillations of $\delta$~Scuti-type
can help constrain the mass and other global parameters of a star, and
the high-order $g$-modes can be used to probe the deep interiors. One
star with such a hybrid pattern of oscillations, HD~209295, was discovered
very recently (Handler 1999b).

The theoretical $\delta$~Scuti instability domains in different diagrams
can be used to make detailed comparisons with existing and future observed
data on these stars.


\acknowledgments

The author thanks B.~Paczy\'nski, M.~Koz{\l}owski and R.~Sienkiewicz
for the stellar evolution code, M.~Jerzykiewicz and R.~Napiwotzki for
the program UVBYBETA which transforms stellar photometric data into
theoretical parameters, and M.~Breger, J.~Christensen-Dalsgaard,
W.~A.~Dziembowski and M.~H.~Montgomery for helpful discussions and comments.
This work has been supported in part by
the grants RFBR--98--02--16734 and KBN--2--P03D--014--14.  Part of the
investigation has been supported by the Austrian Fonds zur F\"{o}rderung
der wissenschaftlichen Forschung, project number S7304.


\end{document}